\documentclass[12pt]{article}

\usepackage{amsfonts}
\usepackage[mathscr]{eucal}
\usepackage{bm}
\usepackage{cite}

\newcommand{\be}{\begin{equation}}
\newcommand{\ee}{\end{equation}}
\newcommand{\bea}{\begin{eqnarray}}
\newcommand{\eea}{\end{eqnarray}}

\newcommand{\p}[1]{(\ref{#1})}

\newcommand{\+}{\dagger}

\begin{document}

\begin{titlepage}

\vspace*{1.5cm}

\renewcommand{\thefootnote}{\dag}
\begin{center}

{\LARGE\bf Relativistic generalization of}

\vspace{0.45cm}

{\LARGE\bf the rational Calogero model}

\vspace{2.5cm}

{\large\bf Sergey Fedoruk }
 \vspace{1.5cm}

{\it Bogoliubov Laboratory of Theoretical Physics, }\\
{\it Joint Institute for Nuclear Research,}\\
{\it 141980 Dubna, Moscow region, Russia} \\
\vspace{0.1cm}

{\tt fedoruk@theor.jinr.ru}

\vspace{1.5cm}

\end{center}

\vspace{2.0cm} \vskip 0.6truecm \nopagebreak

\begin{abstract}
\noindent
\qquad
A relativistic generalization of the rational Calogero model is obtained by using the deformation of a gauging matrix system with extra semi-dynamical variables. The Hamiltonian of this system is derived by imposing the gauge fixing conditions and eliminating gauge degrees of freedom. The integrability of the proposed relativistic model is proved.
\end{abstract}

\vspace{3cm}

\newpage

\end{titlepage}

\setcounter{footnote}{0}
\setcounter{equation}0
\section{Introduction}

In \cite{FIL08,superc}, a supersymmetric generalization of the many-particle rational Calogero systems \cite{C}
was constructed (see \cite{OP,Poly-rev,Ar} for reviews) by using the supersymmetric gauging procedure of the matrix system \cite{DI-06-1}.
In the pure bosonic case, the gauging matrix description of the Calogero models was considered in  \cite{Poly-gauge,Gorsky}.

Recently, there has been great interest in generalizing the Calogero models, in particular, its extension to the relativistic case. A well-known generalization of this type is the Ruijsenaars-Schneider models \cite{RuSch}
(see also the description of these models in \cite{BrSas,FGorNR,GorMir,Ar} and references therein).
However, the generalization of a model like this to the supersymmetric case causes certain problems at the present time (see \cite{ShSu,BTW,Gal,KL,KKL}).
One explanation for this is the lack of a suitable description of the Lagrangian of
the Ruijsenaars-Schneider model.\footnote{A rather complex Lagrangian in \cite{BrSas} obtained from the Ruijsenaars-Schneider Hamiltonian by using the Legendre transform is not well suited for the supersymmetrization procedure.}

However, as noted above, one of the most efficient ways to find the Lagrangian formalism for many-particle Calogero-like systems is to use the gauging procedure for suitable dynamical matrix systems.
For this reason, in this short article, we consider obtaining a relativistic generalization of the Calogero model precisely with the help of
a gauging matrix system.
In this case, we will use natural relativistic relations from elementary particle physics and will not impose a strict requirement that the found relativistic generalization of the Calogero model and the Ruijsenaars-Schneider model coincide.

In the gauging approach, the Calogero system is described by the Lagrangian \cite{Poly-gauge,Gorsky,FIL08,superc}
\begin{equation}
\label{Cal-0}
L_{\rm C}    =  \frac{m}2\ {\rm Tr}\Big( \nabla\! X \nabla\! X\Big) + \kappa\, {\rm Tr}A
\ + \ \frac{i}{2}\, \Big(\bar Z \nabla\! Z - \nabla\! \bar Z Z\Big)
\,,
\end{equation}
where
\begin{equation}
\label{nabla-0}
\nabla\! X = \dot X +i\, [A, X]\,,\qquad \nabla\! Z = \dot Z + iAZ\,, \quad \nabla\! \bar Z = \dot {\bar Z} - iZ A
\end{equation}
are the covariant derivatives of the positive definite Hermitian $c$-number $(n{\times}n)$-matrix field
$$
X(t):=\|X_a{}^b(t)\|\,, \qquad ({X_a{}^b})^* =X_b{}^a\,, \qquad \det X \neq 0\,,
$$
whereas
$$
Z(t):=\|Z_a(t)\|\,, \qquad \bar Z(t):=\|\bar Z^a(t)\|\,, \qquad \bar Z^a = ({Z_a})^*
$$
is the complex $c$-number $\mathrm{U}(n)$-spinor field.
The index $a$ takes $n$ values:  $a=1,\ldots,n$. The quantity $\kappa$ is a real constant and $m$ is a mass parameter.
The dot in expressions like $\dot X$ denotes the derivative with respect to the time variable $t$.

The Hermitian $c$-number $(n{\times}n)$--matrix gauge field
$$
A(t):=\|A_a{}^b(t)\|\,, \qquad ({A_a{}^b})^* =A_b{}^a\,,
$$
that is present in the Lagrangian \p{Cal-0} and in the definitions of the covariant derivatives \p{nabla-0}
is the gauge field for the $\mathrm{U}(n)$ local invariance
\begin{equation}
\label{gauge-trans}
X \rightarrow \, g X g^\+ \,, \qquad  Z \rightarrow \, g Z \,, \quad \bar Z
\rightarrow \,
\bar Z g^\+\,,
\qquad
A \rightarrow \, g A g^\+ +i \dot g g^\+\,,
\end{equation}
where $g(\tau )\in \mathrm{U}(n)$. Fixing gauge for this symmetry and eliminating auxiliary degrees of freedom, we obtain that the model \p{Cal-0}
describes the $n$-particle rational Calogero model in which $\kappa$ plays the role of a coupling constant.
In this formulation, the Calogero pair interaction between particles arises due to a gauging procedure of the matrix model.

In the free case without gauge interaction, the model \p{Cal-0} describes $n$ non-relativistic particles in $D=1+1$ space-time
(with one space dimension).
The relativization of this simple system is standard. In the one-particle case, instead of the Lagrangian
$\displaystyle L_{nr}=\frac12\, m \dot x\dot x$, it is necessary to take the Lagrangian
$\displaystyle L_{r}=-mc^2 \sqrt{1-\frac{1}{c^2} \,\dot x\dot x}$, where $c$ is the real constant (``speed of light'').
Of course, in the  limit $c\to\infty$ the Lagrangian $L_{r}$ turns into to $L_{nr}$ up to the term $mc^2$.
The Hamiltonian of this one-particle relativistic system has the form
$\displaystyle H_{r}=c \sqrt{m^2c^2+p^2}$ where $p$ is the momentum for the coordinate $x$.
When passing from the phase variables $(x,p)$ to the variables $(q,w)$ determined by the relations
$\displaystyle p=mc\sinh w$, $\displaystyle q=mc\,x\cosh w$, where $w$ is the dimensionless rapidity of the particle,
the relativistic Hamiltonian takes the form $\displaystyle H_{r}=mc^2\cosh w$.

In the case of $n$ non-interacting equal-mass particles described
by the phase variables $(x_a,p_a)$ or $(q_a,w_a)$, $a=1,\ldots,n$, the corresponding Lagrangian and Hamiltonian have the form
\begin{equation}
\label{LH-free}
L_{r}=-mc^2 \sum_a\sqrt{1-\frac{1}{c^2} \,\dot x_a\dot x_a} \,,\qquad
H_{r}=mc^2\sum_a\cosh w_a \,.
\end{equation}
This Lagrangian is represented in the matrix form if we introduce the $(n{\times}n)$ diagonal matrix
$\|X_a{}^b(t)\|$ with the quantities $x_a$ on the diagonal. Then the relativistic and nonrelativistic free Lagrangians are rewritten in the form
\begin{equation}
\label{L-free}
L_{r}=-mc^2\, {\rm Tr}\,\sqrt{1-\frac{1}{c^2} \,\dot X\dot X} \,,\qquad L_{nr}=\frac{m}{2}\,{\rm Tr}\,\big(\dot X\dot X\big)\,.
\end{equation}
This simple fact tells us that
in order to obtain a relativistic generalization of the Calogero model,
we can make a similar change in the matrix system \p{Cal-0}.
That is, as a relativistic generalization of the Calogero system we consider the model \p{Cal-0}
in which the first term is replaced by
\begin{equation}
\label{Cal-1}
-mc^2\, {\rm Tr}\,\sqrt{1-\frac{1}{c^2} \,\nabla\! X\nabla\! X} \,.
\end{equation}
Such a relativistic deformation of the Calogero system, which is natural from a physical point of view,
will be the subject of this paper.

The plan of the paper is as follows.
In Section\,2, the matrix relativistic generalization of the Calogero rational system is presented.
Here we give the Lagrangian of this matrix system with semi-dynamical degrees of freedom and its Hamiltonian with the constrains which generate
$\mathrm{U}(n)$ gauge symmetry.
Section\,3 presents the system considered here after fixing all gauges.
The resulting system is described by $n$ coordinates and $n$ momenta, just like the Calogero model.
But unlike the latter system, the resulting system is its relativistic generalization:
the rational Calogero system is obtained only in the nonrelativistic limit $c{\to}\,\infty$.
Section\,4 is devoted to a discussion of the integrability of the constructed system.
It is shown here that the Lax pair for the rational Calogero system and its integrability can be easily obtained
from the gauging matrix system equivalent to it.
Using this method, the integrability of the relativistic system with a full set of conserved charges is obtained.
Section\,5 contains concluding remarks.

\setcounter{equation}0
\section{Gauging model: Lagrangian, Hamiltonian and constraints}

As noted in the introduction, to describe the relativistic generalization of the Calogero system, we consider the Lagrangian
\begin{equation}
\label{Cal-rel}
L_{\rm rC}    =  -mc^2\, {\rm Tr}\,\sqrt{1-\frac{1}{c^2} \,\nabla\! X\nabla\! X} \ + \ \kappa\, {\rm Tr}A
\ + \ \frac{i}{2}\, \Big(\bar Z \nabla\! Z - \nabla\! \bar Z Z\Big) \,.
\end{equation}
In the limit $c\to\infty$, the following relation
\begin{equation}
\label{sq-lim}
\sqrt{1-\frac{1}{c^2} \,\nabla\! X\nabla\! X} \ \simeq \ 1-\frac{1}{2c^2} \,\nabla\! X\nabla\! X
\end{equation}
holds, and
\begin{equation}
\label{L-lim}
L_{\rm rC} \ \simeq \ L_{\rm C} \ - \ n\,mc^2\,,
\end{equation}
where $L_{\rm C}$ is the Lagrangian \p{Cal-0}, $n$ defines the dimension of the $(n{\times}n)$--matrix $X$
and  $(nmc^2)$ is the numerical term.
Therefore, in the nonrelativistic limit, the system \p{Cal-rel} transforms to the $n$-particle rational Calogero system \p{Cal-0}.

Let us consider the Hamiltonization of the matrix system (\ref{Cal-rel}).

The system with the Lagrangian (\ref{Cal-rel}) is described by pairs of phase variables
$(X_a{}^b, P_c{}^d)$, $(Z_a, \mathcal{P}^b)$, $(\bar Z^a, \bar{\mathcal{P}}_b)$
whose nonzero Poisson brackets have the following form:
\begin{equation}\label{PB-X}
\{X_a{}^b, P_c{}^d \}_{\scriptstyle{\mathrm{P}}} =  \delta_a^d \delta_c^b \,,\qquad
\{Z_a, \mathcal{P}^b \}_{\scriptstyle{\mathrm{P}}} =  \delta_a^b \,,\quad
\{\bar Z^a, \bar{\mathcal{P}}_b \}_{\scriptstyle{\mathrm{P}}} =  \delta_b^a  \,.
\end{equation}
The derivation of $\mathrm{U}(n)$-spinor momenta $\mathcal{P}^a$, $\bar{\mathcal{P}}_a$ yields the primary constraints
\begin{equation}\label{const-Z}
G^a:=\mathcal{P}^a - \frac{i}2 \, \bar Z^a \approx 0\,, \qquad
\bar G_a:=\bar{\mathcal{P}}_a + \frac{i}2 \, Z_a \approx 0 .
\end{equation}
Besides, the matrix momentum of $X_a{}^b$ has the form
\begin{equation}\label{P-X}
P_a{}^b =\frac{\partial L_{\rm rC}}{\partial {\dot{X}}_b{}^a}  =
m\left[\nabla X \left(1-\frac{1}{c^2} \,\nabla\! X\nabla\! X\right)^{-1/2}\right]_a^{\,\,\,\,\,b}
\end{equation}
and the momenta of the coordinates $A_a{}^b$ are zero.

The canonical Hamiltonian of the system is
\begin{equation}\label{t-Ham}
H=\ P_b{}^a \dot X_a{}^b + \mathcal{P}^a \dot Z_a + \bar{\mathcal{P}}_a \dot{\bar Z}^a - L_{\rm rC}\ =\ mc^2\,{\rm Tr}\sqrt{1+\frac{1}{m^2c^2} \,P^2}+ {\rm Tr}\big(A F \big)\,,
\end{equation}
where the second term ${\rm Tr}\big(A F \big)$ uses the matrix
\begin{equation}\label{F-constr}
F_a{}^b := i[P,X]_a{}^b + Z_a\bar Z^b- \kappa\,\delta_a{}^b\,.
\end{equation}
Vanishing momenta of the variables $A_a{}^b$ indicate that
quantities  \p{F-constr} define the secondary constraints
\begin{equation}\label{F-constr1}
F_a{}^b \approx 0
\end{equation}
and $A_a{}^b$ in the Hamiltonian  \p{t-Ham} play the role of Lagrange multipliers for these constraints.

The constraints \p{const-Z} are the second class constraints.
Using the Dirac brackets for them and eliminating $\mathcal{P}$-momenta, we obtain that
the nonvanishing  Dirac brackets of the rest phase variables take the form
\begin{equation}\label{DB-XZ}
\{X_a{}^b, P_c{}^d \}_{\scriptstyle{\mathrm{D}}} =  \delta_a^d \delta_c^b \,,
\qquad
\{Z_a, \bar Z^b \}_{\scriptstyle{\mathrm{D}}} =  -i\delta_a^b \,.
\end{equation}

The remaining constraints $F_a{}^b=(F_b{}^a)^*$, defined in  \p{F-constr1}, form the $u(n)$ algebra with respect to the Dirac brackets \p{DB-XZ}:
\begin{equation}\label{DB-FF}
\{F_a{}^b, F_c{}^d \}_{\scriptstyle{\mathrm{D}}} =-i \delta_a{}^d F_c{}^b + i \delta_c{}^b F_a{}^d \,.
\end{equation}
So the constraints \p{F-constr}, \p{F-constr1} are the first class ones and generate $\mathrm{U}(n)$ transformations
\begin{equation}\label{Un-tran}
\delta X_a{}^b =-i[\alpha,X]_a{}^b \,, \quad \delta P_a{}^b =-i[\alpha,P]_a{}^b\,,\quad
\delta Z_a =-i(\alpha Z)_a \,, \quad
\delta  \bar Z_a =i(\bar Z \alpha)^a\,,
\end{equation}
where $\alpha_a{}^b(\tau)=(\alpha_b{}^a(\tau))^*$ are the local parameters.

Thus, the resulting relativistic matrix system is described by the Hamiltonian
\begin{equation}\label{tot-Ham}
H \ =\ H_{\rm rC}+ \lambda_b{}^a F_a{}^b\,,
\end{equation}
where the first term has the following form
\begin{equation}
\label{Ham-matrix1}
H_{\rm rC}  \  = \
mc^2\,{\rm Tr}\sqrt{1+\frac{1}{m^2c^2} \,P^2}\,,
\end{equation}
The quantities $\lambda_a{}^b$ in \p{tot-Ham} are the Lagrange multipliers of the constraints \p{F-constr1},
and the phase-space functions $F_a{}^b$ are defined in \p{F-constr}.

\setcounter{equation}0
\section{Gauge-fixing in the matrix system}

The gauges $X_a{}^b\,{=}\,0$ at $a\,{\neq}\,b$ fix the local transformations \p{Un-tran} with the parameters $\alpha_a{}^b(\tau)$, $a{\neq}b$
generated by the off-diagonal constraints $F_a{}^b \,{\approx}\,0$, $a\,{\neq}\,b$ in the set  \p{F-constr}, \p{F-constr1}.
So, similarly to  \cite{FIL08}, we take the gauge fixing in the form
\begin{equation}\label{x-fix}
x_a{}^b\approx 0 \,,
\end{equation}
where $x_a{}^b$ and $p_a{}^b$, $a\,{\neq}\,b$ are the off-diagonal parts in the matrix expansions
\begin{equation}\label{XP-exp}
X_a{}^b =x_a \delta_a{}^b + x_a{}^b\,,
\qquad
P_a{}^b = p_a \delta_a{}^b + p_a{}^b\,.
\end{equation}
In addition, using
the constraints $F_a{}^b\,{\approx}\,0$, $a\,{\neq}\,b$, we express the momenta $p_a{}^b$ through the remaining  phase variables:
\begin{equation}\label{p-exp}
p_a{}^b= -\frac{i\,Z_a\bar Z^b}{x_a-x_b}\,,\qquad  a\,{\neq}\,b\,.
\end{equation}
Thus, the partial gauge fixing conditions \p{x-fix} and the constraints \p{p-exp} remove $2n(n-1)$ phase variables
$x_a{}^b$ and $p_a{}^b$, $a\,{\neq}\,b$.

Thus, after partial gauge fixing, the phase space of the considered system is defined
by $2n$ real variables $x_a$, $p_a$ and $n$ complex variables $Z_a$.
Due to the ``resolved form'' of gauge fixing conditions \p{x-fix} with respect to the removed variables $x_a{}^b$,
the Dirac brackets for the other phase variables remain unchanged.  Their nonvanishing Dirac brackets are
\begin{equation}\label{DB-xp-z}
\{x_a, p_b \}^{'}_{\scriptstyle{\mathrm{D}}} =  \delta_{ab} \,,
\qquad \{Z_a, \bar Z^b \}^{'}_{\scriptstyle{\mathrm{D}}} =  -i\,\delta_a^b \,.
\end{equation}

The remaining phase variables $x_a$, $p_a$, $Z_a$, $\bar Z^a$ are subject to $n$ residual first class constraints (diagonal parts of \p{F-constr}):
\begin{equation}\label{F-constr-d}
F_a := Z_a\bar Z^a- \kappa \approx 0 \qquad \mbox{(no summation with respect $a$)}\,,
\end{equation}
where $a$ takes $n$ values, $a=1,\ldots,n$. These constraints \p{F-constr-d} generate the following phase transformations of the $\mathrm{U}(n)$-spinor variables $Z_a$:
\begin{equation}\label{Z-tr}
Z_a \ \to \ e^{i\phi_a}Z_a\,, \qquad \bar Z^a \ \to \ e^{-i\phi_a}\bar Z^a\,.
\end{equation}
This gauge symmetry is fixed by the condition that the variables $Z_a$ are real:
\begin{equation}\label{Z-f}
Z_a = \bar Z^a
\end{equation}
at all values of $a$.  As a result, all variables $Z_a$ are gauge cleaned from the system.
Like the previous gauge fixing, the gauge fixing conditions \p{Z-f} (as well as the constraints \p{F-constr-d}) do not contain the rest variables. Therefore, the introduction of the Dirac bracket for the conditions \p{F-constr-d} and \p{Z-f}
does not change the commutation relations for the remaining variables $x_a$ and $p_a$.
In this case, the equations
\begin{equation}\label{Z2-constr}
Z_a\bar Z^a=\kappa \qquad \forall \ a \qquad \mbox{(no summation with respect to $a$)}
\end{equation}
hold in the strong sense.

In the gauges \p{x-fix}, \p{p-exp}, \p{Z2-constr},
the matrix momentum $P_a{}^b$  is equal to the matrix
\begin{equation}\label{P-fix}
\tilde P_a{}^b = p_a\,\delta_a{}^b
\ - \ i (1-\delta_a{}^b) \frac{\kappa}{(x_a-x_b)}  \,,
\end{equation}
depending on the remaining variables $x_a$, $p_a$.
In this case, the Hamiltonian of the system \p{Ham-matrix1} takes the form
\begin{equation}\label{Ham-fix}
\mathrm{H}_{\rm rC} \ = \ c \,{\rm Tr}\sqrt{m^2c^2+\tilde P^2}
\,,
\end{equation}
where the matrix $\tilde P^2$ obtained from the matrix \p{P-fix} is equal to
\begin{eqnarray}\label{P2-fix}
(\tilde P^2)_a{}^b & = & \left[(p_a)^2+\sum_{c\neq a}\frac{\kappa^2}{(x_a-x_c)^2} \right]\delta_a{}^b \\
&& +\, (1-\delta_a{}^b) \left[ -i\,\frac{p_a+p_b}{x_a-x_b} +
\sum_{c{\neq}a,c{\neq}b}\frac{\kappa}{(x_a-x_c)(x_b-x_c)}  \right] Z_a\bar Z^b \,. \nonumber
\end{eqnarray}

In the nonrelativistic limit $c\to\infty$, the Hamiltonian \p{Ham-fix} takes the form
\begin{equation}\label{Ham-fix-lim}
\mathrm{H}_{\rm rC}  =  mc^2 \,{\rm Tr}\sqrt{1+\frac{1}{m^2c^2}\,\tilde P^2} \ \simeq \
\frac{1}{2m}\,{\rm Tr}\left(\tilde P^2\right) \ + \ nmc^2\,.
\end{equation}
Therefore, up to a numerical term in the nonrelativistic limit $c\to\infty$ the Hamiltonian \p{Ham-fix}
of the relativistic system considered here becomes
\begin{equation}\label{Ham-fix-lim-nr}
\mathrm{H}_{\rm rC} \ \simeq \ \mathrm{H}_{\rm C} \ + \ n\,mc^2\,,
\end{equation}
where
\begin{equation}\label{Ham-Cal}
\mathrm{H}_{\rm C} \  =  \ \frac{1}{2m}\sum_a (p_a)^2 \ + \ \sum_{a>b}\frac{\kappa^2/m}{(x_a-x_b)^2}
\end{equation}
is the Hamiltonian of the Calogero system.
Thus, the system considered here is the relativistic generalization (or the relativistic deformation)
of the Calogero rational system.

The removal of the radical in the expression of the Hamiltonian \p{Ham-fix}
is performed in the standard way by diagonalizing the matrix $m^2c^2+\tilde P^2$ inside the square root.
Since this matrix is the positive-definite Hermitian matrix,
the diagonal matrix $\Lambda_a{}^b=\Lambda_a \delta_a{}^b$ is obtained by using the unitary transformation:
\begin{equation}\label{trans-diag}
m^2c^2+\tilde P^2 \  =  \ U\Lambda U^\dagger\,,
\end{equation}
where $U\in \mathrm{SU}(n)$.
After deriving the expressions $\Lambda_a=\Lambda_a(x_b,p_b)$, the Hamiltonian \p{Ham-fix} takes the form
\begin{equation}\label{Ham-fix-ev}
\mathrm{H}_{\rm rC} \ = \ c \,\sum_a \sqrt{\Lambda_a(x_b,p_b)}
\,.
\end{equation}

The eigenvalues $\Lambda_a$ themselves are defined as solutions of the characteristic equation
\begin{equation}\label{ch-eq-P2}
\det\left(m^2c^2+\tilde P^2 - \Lambda \mathbf{1}_n\right)=  \prod_a\left(\Lambda_a-\Lambda \right)=0\,.
\end{equation}
Since the matrix $m^2c^2+\tilde P^2$ is a polynomial function of the matrix $\tilde P$,
the eigenvalues $\Lambda_a$ are determined by the same function (see for example \cite{FadS})
\begin{equation}\label{Lambda-lambda}
\Lambda_a=m^2c^2+(\lambda_a)^2
\end{equation}
of the eigenvalues $\lambda_a(x_b,p_b)$ of the matrix $\tilde P$:
\begin{equation}\label{ch-eq-P1}
\det\left(\tilde P - \lambda \mathbf{1}_n\right)=  \prod_a\left(\lambda_a-\lambda \right)=0\,.
\end{equation}

In the $n=2$ case, the eigenvalues \p{ch-eq-P1} of the matrix $\tilde P$ equal
\begin{equation}\label{lambda-2}
\lambda_\pm= \frac{p_1+p_2}{2} \ \pm \ \frac12\sqrt{(p_1-p_2)^2+\frac{4\kappa^2}{(x_1-x_2)^2}}\,,
\end{equation}
whereas the eigenvalues \p{ch-eq-P2} take the form
\begin{equation}\label{Lambda-2}
\Lambda_\pm = m^2c^2 \ + \ \frac{(p_1)^2+(p_2)^2}{2}\ + \ \frac{\kappa^2}{(x_1-x_2)^2} \ \pm \ \frac{p_1+p_2}{2}\sqrt{(p_1-p_2)^2+\frac{4\kappa^2}{(x_1-x_2)^2}}\,.
\end{equation}
As a result, in the two-particle case, our relativistic Hamiltonian \p{Ham-fix-ev} has the form
\begin{equation}\label{Ham-fix-ev-2}
\mathrm{H}^{n=2}_{\rm rC} \ = \ c \,\sqrt{\Lambda_+(x_1,x_2,p_1,p_2)} \ + \ c \,\sqrt{\Lambda_-(x_1,x_2,p_1,p_2)}\,.
\end{equation}
where $\Lambda_\pm$ are defined in \p{Lambda-2}.

For $n\,{>}2$, the procedure for finding the eigenvalues  $\lambda_a(x_b,p_b)$ and the Hamiltonian \p{Ham-fix-ev} is similar,
although we cannot present analytic expressions for them in the general case because of the complexity of the $n$-th order characteristic equation \p{ch-eq-P1}.
But even for $n{=}\,2$, Hamiltonian \p{Ham-fix-ev-2} is different from the two-particle Ruijsenaars-Schneider Hamiltonian \cite{RuSch}.
At least, we cannot identify them by using a canonical transformation.
In our opinion, the resulting relativistic model with the Hamiltonian \p{Ham-fix-ev}
is different from the Ruijsenaars-Schneider model \cite{RuSch}.

\setcounter{equation}{0}
\section{Integrability of the relativistic model}

The use of the gauging matrix system as the initial one
and the application of the ``resolved'' gauge-fixing conditions
allows us to conclude about the integrability of the resulting relativistic system.

To illustrate and clarify this procedure, let us first consider the nonrelativistic case, that is,
the gauging formulation of the rational Calogero system \cite{Poly-gauge,Gorsky,FIL08,superc}, which is described by the Lagrangian \p{Cal-0}.

After taking into account the second class constraints \p{const-Z}, the total Hamiltonian of the system \p{Cal-0} is represented by the expression
\begin{equation}\label{t-Ham-0}
H_{\rm T}\ =\ \tilde H_{\rm C} + \tilde\lambda_b{}^a F _a{}^b\,,
\end{equation}
where the first term has the form
\begin{equation}
\label{Ham-matrix-0}
\tilde H_{\rm C}  \  = \ \frac{1}{2m}\,{\rm Tr} \left(P^2\right)\,.
\end{equation}
The quantities $F_a{}^b$ have the form \p{F-constr} and define the constraints \p{F-constr1},
whereas $\tilde\lambda_a{}^b$ are the Lagrange multipliers for them.
The evolution of any quantity $K$ is determined by the Dirac brackets of the total Hamiltonian \p{Ham-matrix-0} with it:
\begin{equation}\label{eq-mo}
\dot K\ =\ \{K,H_{\rm T}\}_{\scriptstyle{\mathrm{D}}}\,.
\end{equation}
Here the Dirac brackets \p{DB-XZ} are used. In particular, the evolution of the matrix momentum $P_a{}^b$ is represented by the commutator
\begin{equation}\label{ev-P}
\dot P_a{}^b\ =\ i\, [\,P,\tilde\lambda\,]_a{}^b\,.
\end{equation}
Therefore, the trace from the $\mathrm{k}$-th power of the matrix $P$
\begin{equation}
\label{I-0}
I_{\mathrm{k}}  \  := \ {\rm Tr} \left(P^{\mathrm{k}}\right),
\end{equation}
is conserved. Thus, the conservation of $n$ charges $I_{\mathrm{k}}$, $\mathrm{k}=1,\ldots,n$
guarantees the integrability of the matrix system \p{Cal-0}.
As we see, this proof of integrability is quite simple in the gauging matrix formulation.

A more standard way of proving integrability of the Calogero system is to consider the Hamiltonian \p{Ham-Cal}
and construct the Lax pair for this system.
But the system with the Hamiltonian \p{Ham-Cal} is physically equivalent to the system with the Lagrangian \p{Cal-0}
and the Hamiltonian \p{t-Ham-0}.
More precisely, the system \p{Cal-0} reproduces the system \p{Ham-Cal} after imposing the gauge-fixing conditions \p{x-fix} and \p{Z-f}.
Due to the introduction of the Dirac bracket, the constraints $F_a{}^b\approx0$ and the gauge-fixing conditions \p{x-fix} and \p{Z-f} are zero in the strong sense and express the variables $Z_a$, $\bar Z^a $, $x_a{}^b$, $p_a{}^b$, $a{\neq}b$ in terms of the variables $x_a$ and $p_a$.
Due to the resolved form of the gauge-fixing conditions \p{x-fix} and \p{Z-f}, the Dirac brackets of the remaining variables remain canonical:
$\{x_a,p_b\}_{\scriptstyle{\mathrm{D}}}=\delta_{ab}$.
In addition, the time-conservation of the gauge-fixing conditions \p{x-fix} and \p{Z-f}
\begin{equation}\label{lambda-eq}
\{x_a{}^b,H_{\rm T}\}_{\scriptstyle{\mathrm{D}}}=0\,,\qquad
\{Z_a-\bar Z^a,H_{\rm T}\}_{\scriptstyle{\mathrm{D}}}=0
\end{equation}
defines the Lagrangian multipliers $\tilde\lambda_a{}^b$ in terms of the remaining phase coordinates:
\begin{equation}\label{lambda-def}
\tilde\lambda_a{}^b \ = \ - \delta_a{}^b \sum_{c\neq a} \frac{\kappa}{m(x_a-x_c)^2} \ + \ (1-\delta_a{}^b)\,\frac{\kappa}{m(x_a-x_b)^2}\,.
\end{equation}

Under imposed gauges \p{x-fix} and \p{Z-f} the momentum matrix $P_a{}^b$ takes the form (see also \p{P-fix})
\begin{equation}\label{P-exp-fix}
\tilde P_a{}^b \ = \ p_a \delta_a{}^b \ - \ i(1-\delta_a{}^b)\,\frac{\kappa}{x_a-x_b}\,,
\end{equation}
which is just the matrix $L$ in the Lax equation
for the Calogero rational system (see for example \cite{OP,Poly-rev})
\begin{equation}\label{Lax-eq}
\dot L = i \,[\,L,M\,]\,.
\end{equation}
Moreover, the matrix $\tilde\lambda_a{}^b$
obtained in \p{lambda-def} coincides with the commonly used matrix $M_a{}^b$ in the Lax equation \p{Lax-eq}
(see for example \cite{OP,Poly-rev}).
That is, the Lax equation  \p{Lax-eq} of the Calogero system is already encoded in the equation of motion \p{ev-P} of the matrix variable in the gauging matrix formulation.

In the relativistic case, the Hamiltonian \p{tot-Ham}, \p{Ham-matrix1} in the matrix formulation is the $c$-deformation of the Calogero Hamiltonian \p{t-Ham-0}, \p{Ham-matrix-0}.
However, the matrix momenta $P_a{}^b$ have the equations of motion similar to equations \p{ev-P}.
Therefore, the quantities $I_{\mathrm{k}}$, $\mathrm{k}=1,\ldots,n$ defined in  \p{I-0} are conserved in the case of the model \p{tot-Ham} \p{Ham-matrix1} like the Hamiltonian itself.
As a result, we obtain the integrability of the relativistic model considered here.

The proof of this statement for the system with the Hamiltonian \p{Ham-fix},
which is obtained from the system \p{tot-Ham}, \p{Ham-matrix1} after imposing the gauge fixing conditions \p{x-fix} and \p{Z-f}, is rather cumbersome,
in contrast to the usual Calogero system.

Due to the complicated dependence of the Hamiltonian function \p{Ham-matrix1},  \p{Ham-fix-ev}
on the phase variables $x_a$ and $p_a$, the equations of motion of the latter have a cumbersome form.
Similar difficulties arise in finding explicit expressions for the Lagrange multipliers $\lambda_a{}^b$ from equations \p{lambda-eq}.
But in the limit $c\to \infty$,
the quantities $\lambda_a{}^b$ are equal to the right-hand sides of expressions \p{lambda-def}:
$\displaystyle\lim_{c\to \infty}\lambda_a{}^b=\tilde\lambda_a{}^b$.
Since $\lambda_a{}^b$ define the matrix $M$ in the Lax equation \p{Lax-eq},
this limit will also be the case for the matrix $M$ of the relativistic generalization \p{Cal-rel}
of the Calogero system.
Thus, we arrive at a natural conclusion:
the Lax equations of the relativistic system \p{Cal-rel} are the $c$-deformation of the Lax equations of the nonrelativistic system
which is the rational Calogero system.

\setcounter{equation}{0}
\section{Concluding remarks}

In this paper, a relativistic generalization of the rational Calogero system is derived.
This relativistic generalization is described by the matrix system with gauge symmetry and additional semi-dynamic variables.
The simplest model of this type describes the many-particle Calogero system, as shown in \cite{Poly-gauge,Gorsky,FIL08,superc}.
The more general system considered here contains the speed-dimension parameter $c$ and
describes the relativistic system that reproduces the Calogero model in the limit $c{\to}\,\infty$.

In the matrix formulation, the relativistic model under consideration is described
by the Hamiltonian \p{Ham-matrix1} and the first class constraints \p{F-constr}, \p{F-constr1}
that generate the $\mathrm{U}(n)$ gauge symmetries.
After gauge fixing and eliminating the auxiliary degrees of freedom,
the physically-equivalent reduced system is described by the Hamiltonian \p{Ham-fix} in the $2n$-dimensional phase space.
The model presented here is different from the Ruijsenaars-Schneider model \cite{RuSch},
which is also a relativistic generalization of the Calogero model.
However, the relativistic system described by the Hamiltonian \p{Ham-fix} is also integrable.
It is difficult to prove integrability for the Hamiltonian \p{Ham-fix},
but it is easy to show it for the Hamiltonian \p{Ham-matrix1} of the equivalent matrix system.

Note that the model presented here, with the Hamiltonian \p{tot-Ham}, \p{Ham-matrix1},
is one of the systems
in which the first term of the Hamiltonian (the part without constraints and Lagrange multipliers) depends only on the matrix momentum $P$.
Although the question of constructing the Lagrangian of this kind of models in the general case remains open,
these systems are all integrable, which can be shown within the framework of the consideration carried out in Sect.\,4.

In the following publications, a supersymmetric generalization of the matrix system \p{Cal-rel} is planned
within the framework of the procedure developed in  \cite{FIL08,superc}.

One more interesting problem is to construct similar relativistic $c$-deformations of other integrable Calogero-like systems,
in particular, the matrix formulation of the hyperbolic Calogero system \cite{FIL19,Fed20}.

\smallskip
\section*{Acknowledgements}
I  would  like  to  thank Gleb Arutyunov, Alexei Isaev, Evgeny Ivanov, Sergey Krivonos, Andrei Mironov, Armen Nersessian
for useful discussions and comments.
This work was supported by the Russian Foundation for Basic Research, grant No.\,20-52-12003.

\end{document}